\documentclass[a4paper,british,english]{IEEEtran}
\usepackage[T1]{fontenc}
\usepackage[latin9]{inputenc}
\usepackage{amsmath}
\usepackage{amssymb}

\makeatletter

\pdfpageheight\paperheight
\pdfpagewidth\paperwidth

\usepackage{fancyhdr}
\usepackage{amsmath,graphicx}

\usepackage{graphicx}

\usepackage{epstopdf}

\usepackage{tikz}
\usetikzlibrary{matrix,shapes,arrows,positioning,chains}

\usepackage{graphicx}
\usetikzlibrary{ arrows,  shapes.misc,  shapes.arrows,  chains,  matrix,  positioning,  scopes,  decorations.pathmorphing,  shadows, calc}
\tikzset{
nonterminal/.style={rectangle, minimum size=6mm, very thick, draw=red!50!black!20, top color=white, bottom color=red!50!black!50, font=\itshape },
terminal/.style={rounded rectangle,  minimum size=6mm, very thick,draw=black!50, top color=white, bottom color=black!20, font=\ttfamily},
junction/.style={circle, draw }}
\pgfsetlinewidth{.4mm}

\usetikzlibrary{decorations.pathmorphing} 
\usetikzlibrary{matrix} 
\usetikzlibrary{arrows} 
\usetikzlibrary{calc} 

\tikzstyle{block} = [draw,rectangle,thick,minimum height=2em,minimum width=2em]
\tikzstyle{sum} = [draw,circle,inner sep=0mm,minimum size=2mm]
\tikzstyle{connector} = [->,thick]
\tikzstyle{line} = [thick]
\tikzstyle{branch} = [circle,inner sep=0pt,minimum size=1mm,fill=black,draw=black]
\tikzstyle{branch2} = [circle,inner sep=0pt,minimum size=0mm,fill=black,draw=black]
\tikzstyle{guide} = []
\tikzstyle{snakeline} = [connector, decorate, decoration={pre length=0.2cm,
                         post length=0.2cm, snake, amplitude=.4mm,
                         segment length=2mm},thick, magenta, ->]

\tikzset{
    state/.style={
           rectangle,
           rounded corners,
           draw=black, very thick,
           minimum height=1em,
           inner sep=1pt,
           text centered,
           },
}

\usetikzlibrary{calc,backgrounds}
\usepackage{verbatim}

\usetikzlibrary{positioning}
\usetikzlibrary{decorations.text}
\usetikzlibrary{decorations.pathmorphing}

\usepackage{bm}

\usepackage{ctable}

\usetikzlibrary{fit}					
\usetikzlibrary{backgrounds}	

\pdfoutput=1

\usepackage{url}

\usepackage{hyperref}

\makeatother

\usepackage{babel}
\begin{document}

\title{O\MakeLowercase{n} S\MakeLowercase{ingle}-C\MakeLowercase{hannel}
S\MakeLowercase{peech} E\MakeLowercase{nhancement} \MakeLowercase{and}
O\MakeLowercase{n} N\MakeLowercase{on}-L\MakeLowercase{inear} M\MakeLowercase{odulation}-D\MakeLowercase{omain}
K\MakeLowercase{alman} F\MakeLowercase{iltering}}

\author{Nikolaos Dionelis, \url{https://www.commsp.ee.ic.ac.uk/~sap/people-nikolaos-dionelis/} \\nikolaos.dionelis11@imperial.ac.uk, \url{http://www.imperial.ac.uk/people/nikolaos.dionelis11}
\\Department of Electrical and Electronic Engineering, Imperial College
London (ICL), London, UK}
\maketitle
\begin{abstract}
This report focuses on algorithms that perform single-channel speech
enhancement. The author of this report uses modulation-domain Kalman
filtering algorithms for speech enhancement, i.e. noise suppression
and dereverberation, in \cite{a367}, \cite{a268}, \cite{a225},
\cite{a360} and \cite{a236}. Modulation-domain Kalman filtering
can be applied for both noise and late reverberation suppression and
in \cite{a268}, \cite{a367}, \cite{a225} and \cite{a360}, various
model-based speech enhancement algorithms that perform modulation-domain
Kalman filtering are designed, implemented and tested. The model-based
enhancement algorithm in \cite{a268} estimates and tracks the speech
phase. The short-time-Fourier-transform-based enhancement algorithm
in \cite{a236} uses the active speech level estimator presented in
\cite{a102}. This report describes how different algorithms perform
speech enhancement and the algorithms discussed in this report are
addressed to researchers interested in monaural speech enhancement.
The algorithms are composed of different processing blocks and techniques
\cite{a172}; understanding the implementation choices made during
the system design is important because this provides insights that
can assist the development of new algorithms.
\end{abstract}


\IEEEpeerreviewmaketitle{
\begin{table}[b]

\IEEEpeerreviewmaketitle{\quad Nikolaos Dionelis is a PhD researcher at Imperial College London (ICL) under the supervision of Mike Brookes (mike.brookes@imperial.ac.uk) during the course of this work. This document has more than 100 references.}
\end{table}
}

\section{Introduction }

Technology is ever evolving with tremendous haste and the demand for
speech enhancement systems is evident. The need for speech enhancement
for human listeners is apparent due to the increase in the number
of smartphone users. Speech enhancement for listeners is also needed
in hearing aids. The requirements for speech enhancement for human
listeners are not the same as for automatic speech recognition (ASR);
nevertheless, the algorithms that perform speech enhancement for human
listeners can be used for ASR. Examples that advocate the latter argument
can be found in the REVERB challenge \cite{a261}, \cite{a262}. In
\cite{a378}, speech enhancement is presented as front-end ASR. Nowadays,
many technology-based applications need speech enhancement as a front-end
system \cite{a378}. For example, ASR algorithms for robot audition
can benefit from the use of speech enhancement as a front-end system.
Smartphone applications also need speech enhancement as a front-end
system. To answer to one's questions, digital assistants such as ``Google
Home'' \cite{a267} and Amazon's ``Alexa'' can also benefit from
the use of front-end speech enhancement. Front-end adaptive dereverberation
has been used in \cite{a267} and \cite{a379}.

Single-channel speech enhancement is different from multi-channel
speech enhancement. Multi-channel speech enhancement can take advantage
of the correlation between the different microphone signals and of
the spatial cues that are related to the configuration of the microphones
\cite{a344} \cite{a345}. Multi-channel speech enhancement can be
performed using beamforming followed by single-channel speech enhancement
\cite{a364} \cite{a379}. Beamforming is utilised for spatial discrimination and
is usually followed by single-channel speech enhancement. The problem
of single-channel/monoaural speech enhancement continues to be of
significant interest to the speech community mainly because multi-channel
enhancement can be performed with a beamformer followed by single-channel
enhancement. Considering the enormous increase in the number of smartphone
users, multi-channel (and thus single-channel) enhancement is needed
as front-end in many applications.

The two main causes of speech degradation are additive noise and room
reverberation, as described in, for example, the ACE challenge \cite{a257}.
Speech recordings are degraded by noise and reverberation when captured
using a near-field or far-field distant microphone within a confined
acoustic space. Noise and reverberation have a detrimental impact
on speech quality and speech intelligibility \cite{a367} \cite{a379}.
Providing robustness to speech systems still remains a challenge due
to noise and reverberation. Background noise, which is also known
as ambient noise, can be stationary or non-stationary \cite{a379}.
Noise can have tonal components that may have strong phase correlation
with speech. Reverberation is a convolutive distortion; a room impulse
response (RIR) includes components at both short and long delays resulting
in both coloration \cite{a47} and reverberation and/or echoes. Reverberation
can be quite long with a reverberation time, $T_{60}$, of more than
$900$ ms. Noise is uncorrelated with speech \cite{a367}, early reflections are
strongly correlated with speech and late reverberation is uncorrelated
with speech. Early reverberation is not perceived as separate sound sources and is correlated with clean speech \cite{a379}.

The goal of speech enhancement is to reduce and ideally eliminate
the effects of both additive noise and room reverberation without
distorting the speech signal \cite{a379} \cite{a380}. The aim is to enhance speech with high
levels of noise in situations where noise is sufficiently high so
that the speech quality is damaged \cite{a380} and in situations
where abrupt changes of noise occur. Such situations arise commonly
when the microphone is some distance away from the target speaker
because the acoustic energy that the microphone receives from the
target speaker decreases with the square of the distance whereas the
noise energy typically remains constant. The aim of speech enhancement
is to improve the perceived quality of speech by suppressing noise
and late reveberation \cite{a379}. In particular, we aim to suppress late reverberation
because early reflections are not perceived as separate sound sources
and usually improve the speech quality and intelligibility of
the degraded signal.

\section{Literature Review}

Single-channel speech enhancement can be performed in different domains
\cite{a268}. The ideal domain should be chosen such that (i) good
statistical models of speech and noise exist in this domain, and (ii)
speech and noise are separable in this domain. Speech and noise are
additive in the time domain and therefore in the complex Short Time
Fourier Transform (STFT) domain \cite{a379}. Speech and noise are
not additive in other domains such as the amplitude, power or log-power
spectral domains. The relation between speech and noise becomes incrementally
complicated in the amplitude spectral domain, the power spectral domain,
the log-spectral domain and the cepstral domain. Modeling speech spectral
log-amplitudes as Gaussian distributions leads to good speech modeling
because the logarithmic scale is a good perceptual measure and because
researchers use super-Gaussian distributions that resemble the log-normal,
such as the Gamma distribution \cite{a126}, to model speech in the
amplitude spectral domain. In this context, using the log-normal distribution
in the amplitude spectral domain is equivalent to using the Gaussian
distribution in the log-spectral domain. Speech signals can be modeled
more accurately using super-Gaussian Laplacian distributions than
using Gaussians in terms of the amplitude spectral coefficients \cite{a184},
\cite{a293}.

The research work in \cite{a367} focuses on model-based speech enhancement
aiming towards both noise suppression and dereverberation. Speech
enhancement is performed in the log-spectral time-frequency domain
using a Kalman filter (KF) to model temporal inter-frame correlations
\cite{a268}. The reasons for choosing the log-spectral time-frequency
domain are related to (i) in the previous paragraph: good statistical
models of speech and noise exist in the log-spectral time-frequency
domain \cite{a225}. Speech spectra are well modelled by Gaussians
in the log-spectral domain (and not so well in other domains) \cite{a268},
mean squared errors in the log-spectral domain are a good measure
to use for perceptual speech quality and the non-nonnegative log-spectral
domain is most suitable for infinite-support Gaussian modeling. The
log-spectral domain is used because of the aforementioned reasons
and because the loudness perception of the peripheral human auditory
system is logarithmic.

Regarding (ii) and regarding the extent to which speech and noise
are separable in the log-spectral time-frequency domain, some noise
types are sparse in time and some are sparse in both time and frequency \cite{a379}.
Speech is sparse in both time and frequency. Intermittent noise is
sparse in time and some noise types are fairly sparse in both time
and frequency. In addition, speech and noise are correlated over successive
frames.

Monoaural speech enhancement is most commonly done in a time-frequency
domain because both speech and, in many cases, interfering noise are
relatively sparse in this domain. Speech is sparse in both time and
frequency, intermittent noise is sparse in time and some noise types
are fairly sparse in both time and frequency. A recent paper that
advocates the argument that speech signals are sparse in both time
and frequency is \cite{a275}. The sparse nature of speech spectrograms
is also utilised in the dereverberation algorithm in \cite{a290}.

Speech enhancement can also be performed in the time domain, even though
speech is not sparse in the time domain. Early speech enhancement
was performed in this domain.

Kalman filtering can be performed in the time domain; there is a plethora
of enhancement algorithms that use a KF in the time domain and they
all have originated from \cite{Paliwal1987}. Kalman filtering in
the time domain needs a KF state of a large dimension; for example,
the KF state dimension is $22$ for a $20$ kHz sample rate and $10$,
or even $14$, for an $8$ kHz sample rate. Kalman filtering in the
time domain, \cite{So2010} \cite{Paliwal1987}, is different from
modulation-domain Kalman filtering, \cite{a203} \cite{a233}. Kalman
filtering in the time domain, as performed in \cite{So2010} and in
\cite{Paliwal1987}, operates in the time domain and changes the spectrum,
without explicitly computing the spectrum. In the same way, modulation-domain
Kalman filtering, as performed in \cite{a203} \cite{a233}, operates
in a spectral time-frequency domain and changes the modulation spectrum,
without explicitly computing it.

The model-based speech enhancement algorithms in \cite{a367} and
in \cite{a268}, which estimates and tracks the clean speech phase,
solve the problem of monaural speech enhancement using modulation-domain
Kalman filtering, which refers to imposing temporal constraints on
a spectral time-frequency domain. Three possible domains are the amplitude
spectral domain, the power spectral domain and the log-spectral domain.
Non-linear adaptive modulation-domain Kalman filtering refers to tracking
the clean speech signal in one of the three spectral domains along
with imposing inter-frame constraints \cite{a268}.

Speech is highly structured and it is mainly structured in its inter-frame
component. Speech is a highly self-correlated signal and, by taking
the inter-frame correlation into account, we are able to develop more
sophisticated algorithms with better noise reduction results \cite{a14}.
Speech has prominent temporal dependency which provides rich information
for speech processing and this is why modulation-domain Kalman filtering
can be performed. The speech enhancement algorithms in \cite{a268}
and \cite{a225} model the temporal dynamics of the speech spectral
log-powers, assuming that the STFT spectral log-power of the current
frame is correlated with the STFT spectral log-power of the neighboring
frames. When the algorithms estimate the spectral log-power of the
clean speech in the current frame, they use the STFT spectral log-powers
of the noisy speech not only in the current frame but also in the
previous ones.

Speech enhancement aims to minimize the effects of additive noise
and room reverberation on the quality and intelligibility of the speech
signal. Speech quality is the measure of noise remaining after the
processing on the speech signal and of how pleasant the resulting
speech sounds, while intelligibility refers to the accuracy of understanding
speech. Enhancement algorithms are designed to remove noise and reverberation
with minimum speech distortion \cite{a379}. There is a trade-off
between speech distortion and noise and reverberation suppression.
Enhancement is challenging due to lack of knowledge about both the
speech and the corrupting noise.

Speech enhancement is most commonly performed in a time-frequency
domain that is related to the STFT and thus using STFT bins \cite{a380}.
The main advantage of utilising the (high) frequency resolution of
STFT bins is that perfect reconstruction is possible in the STFT domain.
Different frequency bands, such as Mel-spaced bands and Bark-spaced
bands, can also be used. The Mel-frequency scale is a perceptually
motivated scale that is linear below $1$ kHz and logarithmic above
$1$ kHz. Gammatone time-domain filters can also be used. The STFT
is popular because it can be made to have perfect reconstruction;
however, Mel-bank or Bark-bank or gammatone filters more closely match
the frequency resolution of human hearing \cite{a360}. To reduce the computational
complexity of signal processing algorithms, matching the frequency
resolution of human hearing is important. Human hearing mainly depends
on low and medium frequencies \cite{a184} \cite{a172} and high spectral
resolution is not always needed at high frequencies \cite{a360}.

Gammatone filters are easy-to-implement real-valued filters, usually
of the eighth order, that match human hearing \cite{a292}. One of
the main advantages of gammatone filters is that no frame segmentation
is needed; the signal is in the time domain during the entire processing
and the time-frequency trade-off is not evident. In this way, no artifacts
are created from frame segmentation. The gammatone time-domain filters
transform the signal into bands and then real-valued gains are computed
for each band. One of the main disadvantages of gammatone filters
is that perfect signal reconstruction is not possible.

Speech enhancement can be performed in different time-frequency domains,
such as the complex STFT domain, the amplitude spectral domain and
the power spectral domain. Other possible time-frequency domains are
the log-spectral domain, the cepstral domain and the (spectral) phase
domain \cite{a217} \cite{a11}. Moreover, speech enhancement can
be performed either using the real and the imaginary parts of the
complex STFT domain \cite{a13} \cite{a251} or using the log real
and the log imaginary parts of the complex STFT domain. Most enhancement
algorithms modify only the amplitude of the spectral components and
leave the phase unchanged for three reasons: (i) estimating the phase
reliably is difficult \cite{a380}, (ii) the ear is largely insensitive to phase,
and (iii) the optimum estimate of the clean speech phase is the noisy
phase under reasonable assumptions. The enhancement problem is to
estimate a real-valued time-frequency gain to apply to the noisy signal.

The real-valued time-frequency gain can be applied in STFT bins but
can be calculated in Mel-spaced frequency bands, as in \cite{a179}
\cite{a175}. According to \cite{a179} \cite{a175}, the speech enhancement
algorithms can first estimate and then interpolate the real-valued
gain in Mel-spaced frequency bands to estimate and apply the real-valued
gain in uniformly-spaced STFT bins.

Spectral subtraction in the magnitude spectral domain (or in the power
spectral domain) was one of the most early enhancement techniques.
Furthermore, regarding traditional enhancement algorithms, Minimum
Mean Square Error (MMSE) \cite{Ephraim1984} and Log-MMSE \cite{Ephraim1985}
are two of the most popular model-based enhancement techniques. The
superiority of Log-MMSE over MMSE can be considered as motivation
for using the log-spectral domain and thus for minimizing the error
in the log-spectral domain. Both MMSE and Log-MMSE assume a uniform
speech phase distribution \cite{a172} and, also, use that speech
and noise are additive in the complex STFT domain \cite{a123}.

MMSE and Log-MMSE can be considered as one group of algorithms since
they are variants of time-frequency gain manipulation. In \cite{a68},
a description of the MMSE and Log-MMSE statistical-based noise reduction
algorithms is given. The Log-MMSE estimator is better in terms of
speech quality than the MMSE estimator since it attenuates the noise
power more without introducing much speech distortion \cite{a88}.
According to \cite{Cappe1994}, MMSE estimators using the decision-directed
approach do not introduce musical noise. However, according to listening
experiments, this claim of \cite{Cappe1994} is not actually true.
In MMSE, \cite{Ephraim1984}, the a posteriori SNR is the noisy speech
power divided by the noise power and the a priori SNR is the clean
speech power divided by the noise power. The traditional MMSE approach,
\cite{Ephraim1984}, uses the decision-directed approach to estimate
the a priori SNR from the a posteriori SNR. The traditional Log-MMSE
approach, \cite{Ephraim1985}, uses the log-power domain. In MMSE,
the model assumes that the STFT coefficient of noisy speech is the
sum of two zero-mean complex Gaussian random variables; the STFT coefficients
of clean speech and noise are modeled with a zero-mean complex Gaussian
distribution \cite{a68}. For complex Gaussian random variables, the
magnitude and phase are independent and this is a common assumption
in speech processing algorithms. In addition, the distribution of
the magnitude is Rayleigh and the distribution of the phase is uniform
in $(-\pi, \pi)$; the latter assumption is common in speech enhancement
algorithms. Several variants of the MMSE and Log-MMSE estimators exist;
super-Gaussian models for speech in the amplitude or power spectral
domains have been proposed after the success of Log-MMSE. Alternative
versions of the MMSE are presented, for example, in \cite{a123},
in \cite{a82} and in \cite{a71}.

More recently, researchers have tried to incorporate phase in speech
modeling \cite{a217}, \cite{a11}. The speech phase is not irrelevant,
\cite{a18}, and in low SNR levels, the ear is sensitive to the phase. Incorporating
the phase leads to applying a complex-valued time-frequency gain to
the noisy speech signal in the complex STFT domain. In \cite{a217}
and \cite{a23}, several speech phase estimation algorithms are
discussed, analysed and tested. The speech separation algorithm in \cite{a380} discretises the difference between the noisy and clean speech phases in a non-uniform way and treats the estimation of the difference between the noisy and clean speech phases as a supervised learning classification problem. In \cite{a380}, the (ideal) ratio mask is also discretised.

Regarding speech phase estimation in non-stationary noisy environments, the model-based speech enhancement algorithm presented in \cite{a268} estimates and tracks the clean speech phase. The STFT-based enhancement algorithm in \cite{a268} performs adaptive non-linear Kalman filtering in the log-magnitude spectral domain to track the speech phase in adverse conditions.

Recently, researchers consider the inter-frame correlation of speech.
In traditional speech enhancement, each time-frame was considered
on its own and inter-frame correlation was not explicitly modeled.
In traditional speech enhancement, such as in MMSE or Log-MMSE, the
local SNR estimate (i.e. either the a priori or the a posteriori local
SNR) was smoothed and this is how inter-frame correlation was indirectly
considered; there was no explicit model for the inter-frame correlation
of speech. Nowadays, the inter-frame correlation of speech can be
modeled using the modulation domain. Regarding modulation-domain algorithms,
the relative spectra (RASTA) and Gabor modulation filters have been
used for enhancement \cite{a273} and are popular as pre-processing
front-end methods to ASR. The RASTA filter is a band-pass filter in
the modulation domain that eliminates low and high modulation frequencies
\cite{a273}.

Modulation-domain Kalman filtering \cite{a203} \cite{a233} is different
from the aforementioned modulation filters in the sense that the modulation-domain
Kalman filtering algorithms do not compute the modulation spectrum.
The modulation-domain Kalman filtering algorithms change/alter the
modulation spectrum but they do not explicitly compute the modulation
spectrum. Modulation-domain Kalman filtering considers the inter-frame
correlation of speech in the spectral domain. With modulation-domain
Kalman filtering, temporal constraints are imposed on a specific time-frequency
domain of speech.

The modulation-domain Kalman filtering technique was first presented
in \cite{a203} \cite{a233} in 2010. Enhancement algorithms can benefit
from including a model of the temporal inter-frame correlation of
speech. With modulation-domain Kalman filtering, each time-frame is
not treated independently and temporal constraints are imposed on
a specific time-frequency spectral domain of speech. In \cite{a203}
\cite{a233}, modulation-domain Kalman filtering in the amplitude
spectral domain is performed with a linear normal KF update step;
both the inter-frame speech correlation modeling and the speech tracking
are performed in the amplitude spectral domain with a modulation-domain
Kalman filter in \cite{a203} \cite{a233}. In this context, Gaussian
distributions are used in the amplitude spectral domain in \cite{a203}
\cite{a233}. The algorithm in \cite{a203} \cite{a233} assumes a
linear distortion equation in the time-frequency amplitude spectral
domain and this is why it performs a linear normal KF update step.

Whereas traditional speech enhancement algorithms treat each time-frame
independently, an alternative approach performs filtering in the modulation
domain. The modulation domain models the time correlation of frames.
The modulation domain models the time evolution of the clean STFT
amplitude domain coefficients in every frequency bin. The algorithms
described in \cite{a123} and \cite{a126} use modulation-domain KFs.

The modulation-domain KF is a good low order linear predictor at modeling
the dynamics of slow changes in the modulation domain and produces
enhanced speech that ``has minimal distortion and residual noise'',
according to \cite{a203} \cite{a233}. The modulation-domain KF is
an adaptive MMSE estimator that uses models of the inter-frame changes
of the amplitude spectrum, the power spectrum or the log-spectrum
of speech. Modulation-domain Kalman filtering for tracking both speech
and noise is possible and beneficial according to \cite{a123}. Noise
tracking using a KF can be beneficial for enhancement \cite{a243},
\cite{a192}. Noise tracking is performed in \cite{a243} and subsequently
in \cite{a241}. In the KF update step, the correlation between speech
and noise samples can be estimated, as in \cite{a225}, \cite{a268}
and \cite{a236}.

Modulation-domain Kalman filtering can be performed in the amplitude
spectral domain, in the power spectral domain or in the log-magnitude
spectral domain \cite{a268}. The KF equations are different in each
case. Modulation-domain Kalman filtering in the log-spectral domain,
minimizing the error in the log-power spectral domain, is performed
in \cite{a225}, in \cite{a268} and in \cite{a236}. Many papers,
such as \cite{a132} and \cite{a133}, relate clean speech and noisy
speech in the log-spectral domain. The non-linear log-spectral distortion
equation is used in \cite{a131} and in \cite{a129}.

Time-frequency cells of the signal in the amplitude, power or log-power
spectral domain can be viewed as features. When speech is distorted
by noise and reverberation, the temporal characteristics of the feature
trajectories are distorted and need to be enhanced. Filtering that
removes variations in the signal that are uncharacteristic of speech,
changing according to the underlying environment conditions, has to
be performed.

Modulation-domain Kalman filtering in \cite{a203} \cite{a233} assumes
that speech and noise add in the amplitude spectral domain. Assuming
additivity of speech and noise in the amplitude spectrum is an approximation
assuming a high instantaneous SNR. The spectral amplitude additivity
assumption corrupts the algorithm's mathematical perfection and is
unreasonable in physical meaning despite that it produces reasonable
results.

The phase factor, $\alpha$, is the cosine of the phase difference
between speech and noise \cite{a111}, \cite{a181}. The phase factor
and the additivity in the power or the amplitude spectral domain are
related to the in-phase and the in-quadrature components \cite{a102}.
When speech and noise are in-phase, $\alpha=1$; when speech and noise
are in-quadrature, $\alpha=0$. According to \cite{a110}, the effect
of the phase factor is small when the noise estimates are poor. On
the contrary, when the noise estimates are accurate, the effect of
$\alpha$ is stronger \cite{a110}. It was noted in \cite{a246} that
the power-sum, log-sum and max-model approximations are usually used
in denoising speech enhancement. Both the power-sum and the log-sum
approximations assume $\alpha=0$ and thus that speech and noise are
in-quadrature. The max-model approximation resembles, but is not identical
to, the $\alpha=0$ assumption. We note that the amplitude-sum approximation
is not mentioned in \cite{a246}. In modulation-domain Kalman filtering,
\cite{a203} \cite{a233}, and in nonnegative matrix factorization
(NMF), \cite{a193}, the amplitude-sum approximation that assumes
$\alpha=1$ is usually used.

Modeling the effect of noise as additive in the power spectral domain
assumes $\alpha=0$. According to \cite{a237}, it is well known that
modeling the effect of additive noise as additive in the power spectral
domain is only an approximation, which breaks down at SNRs close to
$0$ dB. Then, the cross term in the power spectrum can no longer
be neglected \cite{a237} \cite{a270}.

The algorithm in \cite{a202} assumes that the phase factor is zero,
$\alpha=0$. In \cite{a202}, equation (3) is the power spectral domain
assuming that $\alpha=0$. In \cite{a202}, the log-power spectrum
notation is used in equations (4)-(5) if we ignore the convolutive
distortion and therefore the distortion due to the microphone type
and the relative position of the talker or speaker.

The log-power spectrum non-linear distortion equation is $y = x + \log(1 + \exp(n-x) + 2 \alpha \exp(0.5(n-x)))$,
where $y$ is the noisy speech log-power, $x$ is the speech log-power
and $n$ is the noise log-power \cite{a110} \cite{a225}. All the variables are
defined in the log-power spectral domain. According to \cite{a131},
the phase factor can also be modelled with the equation: $y=x+\dfrac{1}{\gamma}\log\left(1+\exp\left(\gamma\times(n-x)\right)\right)$,
using $\gamma$ and not $\alpha$.

Speech enhancement in non-stationary noise environments is a challenging
research area. The modulation domain is an often-used representation
in models of the human auditory system; in speech enhancement, the
modulation domain models the temporal inter-frame correlation of frames
rather than treating each frame independently \cite{a203} \cite{a233}.
Enhancement algorithms can benefit from including a model of the inter-frame
correlation of speech and a number of authors have found that the
performance of a speech enhancer can be improved by using a speech
model that imposes temporal structure \cite{a47}, \cite{a185}, \cite{a222}.
Temporal inter-frame speech correlation modelling can be performed
with a KF with a state of low dimension, as in \cite{a203} and \cite{a126}.
The algorithms in \cite{a123} track the time evolution of the clean
STFT amplitude domain coefficients in every frequency bin. In \cite{a174},
speech inter-frame correlation is modeled. Considering KF algorithms,
many papers, such as \cite{a132} \cite{a133} and \cite{a129}, use
the non-linear observation model relating clean and noisy speech in
the log-spectral domain.

Modulation-domain Kalman filtering can be applied for both noise and
late reverberation suppression and this is why this report discusses
both noise reduction and dereverberation.

The modulation domain models the time correlation of frames and does
not treat each time-frame independently. The algorithms in \cite{a123}
track the time evolution of the clean STFT amplitude domain coefficients
in every frequency. Denoising algorithms that operate in the modulation
domain use overlapping modulation frames and use the KF. Considering
KF-related algorithms, many papers, such as \cite{a132} and \cite{a133},
use the observation model relating clean speech and noisy speech in
the log-power spectrum. The non-linear log-spectral distortion equation
is also used in \cite{a129}. In \cite{a174}, the time-frame speech
correlation is modeled and is then followed by NMF.

According to \cite{a203}, \cite{a126}, \cite{a123} and \cite{a225},
temporal inter-frame speech correlation modeling requires the use
of a KF with a state of low dimension. Motivated by the fact that
inter-frame speech correlation modeling requires the use of a KF with
a hidden state of dimension $2$, we claim that a KF with a hidden
state of dimension $3$ can effectively be utilized for both inter-frame
and intra-frame/frequency speech correlation modeling. We use the
KF prediction step for both inter-frame and intra-frame speech correlation
modeling. Autoregressive (AR) modeling is a mathematical technique
that models correlation and any local correlation can be modeled with
the Markov assumption. In this paper, we use both inter-frame and
intra-frame KF prediction steps and claim that the intra-frame KF
prediction step can be used for frequencies around the pitch and harmonics.
AR modeling for intra-frames will model the correlation among neighboring
frequencies around the pitch and harmonics. In this way, we can better
discriminate clean speech from noise in the log-magnitude spectral
domain.

The algorithms in \cite{a123} operate in the modulation domain and
treat every frequency bin on its own. In this paper, as main innovation,
we advance intra-frame correlation modeling based on modulation-domain
Kalman filtering by utilizing both inter-frame and intra-frame KF
prediction steps. We use Kalman filtering in the log-power STFT spectrum.
Log-spectral features are highly correlated: the behaviour of a certain
frequency band is very similar to the behaviour of the adjacent frequency
bands. Therefore, the log-power STFT spectrum is highly suitable for
intra-frame modeling.

The procedure that is followed in algorithms that perform modulation-domain
Kalman filtering is as follows. The first step of the procedure is
to transform the time domain signals into a suitable time-frequency
representation using the STFT. In this step, the algorithm divides
the time domain signal into overlapping frames, obtained by sliding
a window through the signal. These frames are then transformed into
the frequency domain at a suitable resolution using the Fourier transform.
The sliding window is shifted through the signal with a suitable hop
to obtain a sub-sampled time-frequency representation that allows
for perfect reconstruction. These steps constitute the STFT \cite{a14}
\cite{a74}. The short-time spectra are then divided into their magnitude
and phase components. The magnitude of the short-time spectra is usually
considered on its own to separate speech from noise, leaving the phase
of the short-time spectra unaltered. In modulation-domain Kalman filtering
algorithms, adjacent magnitude short-time spectra are referred to
as modulation frames; modulation frames, with a suitable length and
increment, are used for AR modeling.

The modulation domain models the inter-frame correlation of clean
speech and does not consider each time-frame independently. In \cite{a174},
inter-frame speech correlation is modeled and is then followed by
NMF. Inter-frame correlations of speech are considered in several
papers and books by J. Benesty, i.e. \cite{a14}. Section 4 in \cite{a14}
presents linear filters for inter-frame temporal correlation modeling
of speech \cite{a268}.

Nowadays, speech enhancement algorithms can model the inter-frame
correlation of the speech spectrum. Short-term inter-frame relationships
can be created based on the Markov property with the KF. The algorithm
in \cite{a225} uses modulation-domain KFs. The KF framework, which
is described amongst others in \cite{a220}, is convenient in that
it allows for statistically grounded approaches to tracking. Kalman
filtering uses local inter-frame priors due to the temporal dynamics
modeling of the KF prediction. Inter-frame correlation modeling of
speech is performed in \cite{a222} using Markov Random Fields.

Inter-frame and intra-frame speech correlation modeling has been considered
from 1987 in \cite{Hansen1987} and, subsequently, from 1991 in \cite{Hansen1991}.
According to \cite{Hansen1991}, inter-frame constaints are imposed
on speech to reduce frame-to-frame pole jitter. In \cite{a222}, Markov
Random Fields are used for both inter-frame and intra-frame speech
correlation modeling. Regarding intra-frame speech correlation modeling
in voided frames, equation (2.6) in \cite{a222} correlates a specific
harmonic with the previous and next harmonics using the observation
that harmonics are integer multiples of the fundamental frequency
\cite{a24} \cite{a172}.

According to Sec 2.3 in \cite{a47}, assuming independence between
time-frames is uncommon and ``this assumption could be relaxed by
imposing temporal structure to the speech model with a recurrent neural
network (RNN)''. According to \cite{a185}, in speech enhancement
algorithms, the KF can be used to create short-term dependencies due
to the Markov property while RNNs can be utilised to create long-term
dependencies between time-frames. The latter statement may be true
for the examples considered in \cite{a185} but it is not generally
true for the RNN in Sec. 3 in \cite{a185}. According to \cite{a186},
it can be shown that memory either decays or explodes in such RNNs
that do not have long-short term memory (LSTM) and it is thus not
clear that one can do better than KFs and the Markov property.

Speech signals can be considered to be correlated only for short-time
periods. In the STFT time-frequency domain, inter-frame speech correlation
exists due to both the speech characteristics and the STFT framing
overlaps \cite{a268} \cite{a367}.

According to \cite{a250}, ``noise reduction using inter-frame speech
correlation modeling has been addressed partially in \cite{a37},
\cite{a13} and \cite{a192} where, in the KF prediction step of a
noise reduction method based on Kalman filtering, complex-valued prediction
weights are used to exploit the temporal correlation of successive
speech and noise STFT coefficients''. The authors in \cite{a250}
do not discuss modulation-domain Kalman filtering and omit the references
of \cite{a203} \cite{a233} and of \cite{a126} \cite{a123}. In
addition, the authors in \cite{a250} claim that ``algorithms that
perform inter-frame speech correlation modeling assume perfect knowledge
of theoretical inter-frame correlation'', which is not valid since
any prediction errors are encapsulated in the AR residual. Modulation-domain
Kalman filtering algorithms \cite{a123} assume small errors from
AR modeling on the pre-cleaned noisy spectrum but they also compute
the AR residual \cite{a268}.

Kalman filtering is related to using Gaussian distributions; in modulation-domain
Kalman filtering, at every time step, the posterior is computed using
the KF-based local prior that is assumed to follow a Gaussian distribution.
According to \cite{a189}, speech enhancement based on spectral features,
such as the amplitude, power and log-power spectrum, degrades when
the spectral prior does not accurately model the distribution of the
speech spectra and when the speech and the noise/interference have
similar spectral distributions. Regarding the latter case, babble
noise has a speech-shaped spectral distribution \cite{a172}.

The modulation-domain Kalman filtering algorithms in \cite{a203}
\cite{a233} perform a linear KF update step \cite{a376}; on the
contrary, the modulation-domain Kalman filtering algorithms in \cite{a123},
in \cite{a279} and in \cite{a278} perform a non-linear KF update
step. For example, in \cite{a126}, the modulation-domain Kalman filter
performs a non-linear KF update step involving the Gamma distribution;
the linear KF prediction step is performed in the amplitude spectral
domain and then moment matching is used to obtain a Gamma prior so
that the modified non-linear KF update step is performed using the
Gamma distribution.

Modulation-domain Kalman filtering can be related to Bayesian filtering
and particle filtering. The algorithm in \cite{a245} uses particle
filtering to track time-varying harmonic components in noisy speech.
Furthermore, non-linear adaptive Kalman filtering can be related to
state-space modeling, which is used in the algorithm in \cite{a241}
that performs both noise reduction and dereverberation. In Sec. IV.B
in \cite{a241}, the algorithm tracks the noise in a spectral domain
using AR modeling.

Non-linear Kalman filtering can be used along with uncertainty decoding,
\cite{a230} \cite{a231}, in ASR because it estimates the speech
amplitude spectrum and its variance. According to \cite{a231}, uncertainty
decoding is a promising approach for dynamically tackling the distortions
remaining after speech enhancement using posterior distributions instead
of point estimates. The uncertainty is computed either directly in
the ASR feature domain or propagated from the spectral domain to the
feature domain \cite{a231}. With modulation-domain Kalman filtering,
the uncertainty/variance is computed in the spectral domain.

Adaptive modulation-domain Kalman filtering with a non-linear KF update
step can be related to the hidden dynamic model that is discussed
and explained in section 13.6 in \cite{a270}. The non-linear mapping
from the hidden states to the continuous-valued acoustic features
in equation (13.39) in \cite{a270} resembles the KF update step that
non-linearly relates the continuous-valued clean acoustic features
with the continuous-valued noisy acoustic features. In section 13.6
in \cite{a270}, the top-down generative process of the hidden dynamic
model is analysed; the KF can be explained as a top-down process.

Speech enhancement is difficult especially when the noisy speech signal
is only available from a single channel. Although many single-channel
speech algorithms have been proposed that can improve the SNR of the
noisy speech, they also introduce speech distortion and spurious tonal
artefacts known as musical noise. In noisy conditions, the tradeoff
between speech distortion and noise removal is apparent. According
to the literature and to \cite{a184} and \cite{a183}, if the evolution
of noise is slower than the evolution of speech, and thus if noise
is more stationary than speech, then noise can efficiently be estimated
during the speech pauses. On the contrary, if noise is non-stationary,
then it is more difficult to estimate the noise and this results in
speech degradation \cite{a183}. In this research work, coloured noise
is considered. According to the literature and to \cite{a172} and
\cite{a184}, real-world noise is colored and does not affect the
speech signal uniformly over the entire spectrum \cite{a123}.

Common/typical speech enhancement algorithms work on the STFT magnitudes,
on the STFT powers or on the STFT log-powers, leaving the phase unaltered \cite{a379}.
Other speech enhancement approaches alter the phase by considering
the complex STFT domain, the real and imaginary parts of the complex
STFT domain or the log real and log imaginary parts of the complex
STFT domain. Furthermore, according to the literature \cite{a109}
\cite{a270}, some speech enhancement algorithms operate on the cepstrum
and leave the phase unaltered.

Regarding the complex STFT domain, according to \cite{a13}, performing
complex AR modeling produces more accurate results than tracking the
real and imaginary parts separately and there is no correlation in
successive phase samples.

The cepstral domain is a possible speech processing domain. The cepstrum,
which is different from the complex cepstrum \cite{Oppenheim1975},
can be considered as a smoothed version of the log-spectral domain.
On the one hand, the cepstrum is the inverse Fourier transform of
the logarithm of the magnitude of the Fourier transform. On the other
hand, the complex cepstrum is based on both the magnitude and the
phase of the Fourier transform; the complex cepstrum is the inverse
Fourier transform of the complex logarithm, $\log(r \exp(j \theta)) = \log(r) + j \theta$,
of the Fourier transform \cite{Oppenheim1975}. The cepstrum can be
used for enhancement and it is usually used with Mel bands.

According to the literature and to \cite{a109} and \cite{a270},
the front-end speech recognition system is as follows. A discrete
Fourier transform (DFT) is applied after windowing; next, the power
spectrum is computed, Mel-spaced bands are used, the log operator
is used and then a second Fourier transform is performed. The second
Fourier transform is usually a Discrete Cosine Transform (DCT). The
DCT is performed on the Mel-spaced log-spectrum to compute the ceptrsum.
The output of the DCT is approximately decorrelated; hence, the decorrelated
features can be modelled with a Gaussian distribution that has a diagonal
covariance matrix \cite{a109} \cite{a270}. The latter observation
that the decorrelated DCT output features are usually modelled with
a Gaussian distribution that has a diagonal covariance matrix is interesting.
Speech enhancement as a front-end to speech recognition aims to enhance
either the final feature of the cepstrum or any intermediate feature.

The speech enhancement algorithms that work on the STFT magnitudes
try to minimize the error in the amplitude spectral domain. Likewise,
the algorithms that work on the STFT powers try to minimize the error
in the power spectral domain and the algorithms that operate on the
STFT log-powers try to minimize the error in the log-spectral domain.
In this sense, the enhancement algorithms that work on the STFT log-powers
resemble the algorithms that use the log mean squared error (MSE)
spectral distortion metric \cite{a88} \cite{a184}. In \cite{a88},
P. C. Loizou examines the use of perceptual distortion metrics, such
as the Itakura-Saito (IS) distortion and the hyperbolic-cosine (COSH)
distortion, instead of the MSE and the log-MSE. Perceptual distortion
metrics had been used for speech recognition before 2005 and, in 2005
\cite{a88}, perceptual distortion metrics were used for speech enhancement
and for estimating clean speech in the amplitude spectral domain.

Considering the amplitude, power and log-power spectral domain and
the perceptual distortion metrics \cite{a88} \cite{a184}, speech
can be estimated and/or tracked in perceptually motivated time-frequency
domains, such as the IS-spectral domain or the COSH-spectral domain.
Perceptually motivated spectral time-frequency domains have not been
used for speech tracking.

\section{Additional Literature Review}

The non-linear KF algorithm in \cite{a268} is a model-based speech enhancement algorithm based on parametric estimation. KF algorithms are different from
data-driven algorithms, such as \cite{a347} and \cite{a242}. Data-driven
neural network algorithms consider all frequency bins simultaneously
and are different from parametric estimation algorithms that operate
on a per frequency bin basis \cite{a353} \cite{a345}. In \cite{a347},
a LSTM RNN is used to estimate late reverberation that is then subtracted
from the reverberant speech signal to estimate the anechoic dry speech.
Supervised learning is examined in the PhD Theses \cite{a323} and
\cite{a351}.

A novel direction in speech enhancement refers to the use of neural
networks (NNs) and deep NNs \cite{a378} \cite{a379}. NN-based speech enhancement, which has
been examined in \cite{a380}, \cite{a351} and \cite{a323}, can
be used. Amongst other places, deep NNs are mathematically described
and discussed in chapter 4 in \cite{a270}; several examples of NN-based
enhancement algorithms can be found in \cite{a288}, \cite{a193},
\cite{a271} and \cite{a227}. NNs perform frequency intra-frame correlation
modeling since their inputs are the noisy speech in the amplitude
spectral domain, the power spectral domain or the log-spectral domain.
In NNs, inter-frame correlation of speech is modeled by considering
context frames, which can be considered as overlapping modulation
frames, as inputs to the NN. However, this speech inter-frame correlation
modeling often leads to artefacts, decreasing the speech artefact
ratio in source separation, according to slide 35 in \cite{a288}.
Specifically, according to slide 35 in \cite{a288}, frame-by-frame
denoising with NNs produces comparable results to NNs with context
frames in terms of separation metrics.

In contrast to NNs \cite{a380}, model-based enhancement algorithms that perform modulation-domain
Kalman filtering use few parameters and utilise the equations relating
speech and noise in the complex STFT domain. Specific equations relating
speech and noise in the spectral domain are used and the relationship
between speech and noise is not learned from training data. Non-linear
Kalman filtering algorithms model the speech inter-frame
correlation in the STFT domain but not the speech intra-frame correlation
in the STFT domain. NNs are robust to small variations of the training data \cite{a199}
and are sensitive to training techniques and training samples \cite{a101}
\cite{a199}. NNs over-parametrise the speech enhancement problem
and, moreover, NNs assume that training and testing samples are independent and identically
distributed (iid) in most cases.

The preceding paragraphs are not just a discussion of machine-learning
versus model-based techniques, which is a well rehearsed discussion \cite{a379}.
The observation that NNs over-parametrise the problem while modulation-domain
Kalman filtering algorithms use few parameters for each frequency
bin to parametrise the speech enhancement problem is important. The observation
that unseen noise types, unseen SNRs, unseen reverberation times and
other unseen conditions affect the performance of NNs is also significant.
Furthermore, another important observation is that the training of
NNs is based on local minima: training NNs involves non-convex optimization
\cite{a379} and the use of good priors is critical. Good priors can
be considered as regularization, like dropout, to avoid overfitting.
The training procedure has to reach a good local minimum that will
lead to network parameters that will make the NN generalize well to
unseen test data \cite{a101}. During inference, NNs are very fast
and they also require low computation \cite{a288}.

Ideal ratio masks and complex ideal ratio masks usually utilise a
NN to estimate the real and the imaginary parts of the complex STFT
of speech, as discussed in \cite{a283}. Ideal ratio masks compute
a real-valued time-frequency gain; complex ideal ratio masks find
a complex-valued time-frequency gain. Binary masking is different
from ratio masking because it is based on classification and on hard
labels (not soft labels).

Another contemporary direction in speech enhancement refers to the
use of end-to-end systems. End-to-end systems operate in the time
domain and depend on NN training, both on the training data and the training procedure \cite{a379} \cite{a380}.

Regarding dereverberation \cite{a269}, a few KF-based dereverberation
algorithms exist in the literature. Dereverberation aims to remove
echo and reverberation effects from speech signals for improved speech
quality and intelligibility. Reverberation causes smearing across
time and frequency; reverberation tends to spread speech energy over
time. This time-energy spreading has two distinct effects: (i) the
energy in individual phonemes become more spread out in time and,
consequently, plosives have a delayed onset and decay and fricatives
are smoothed, and (ii) preceding phonemes blur into the current phonemes.
According to the literature \cite{a262} \cite{a269}, the effect
of (ii) is most apparent when a vowel precedes a consonant. Both (i)
and (ii) reduce speech quality and speech intelligibility.

Speech captured with a distant microphone inevitably contains both
reverberation and noise. In the time domain, the reverberant noisy
speech signal, $y(t)$, can be expressed as $y(t) = h(t) \ast s(t) + n(t)$
where $h(t)$ is the RIR between the talker and the microphone, $s(t)$
is the clean speech signal, $n(t)$ is the noise signal and $\ast$
is the convolution operator. Most dereverberation algorithms are mostly
concerned with the effects of the late reflections. The temporal masking
properties of the human ear cause the early reflections to reinforce
the direct sound \cite{a269}, and this is why early reverberation
and early reflections enhance the quality of degraded speech signals.

The reverberation time, $T_{60}$, and the Direct to Reverberant energy
ratio (DRR) are the two main parameters of reverberation \cite{a286}
\cite{a379}. The $T_{60}$ quantifies the reverberation duration
along time and is defined as the time interval required for a sound
level to decay $60$ dB after ceasing its original stimulus. The DRR
describes the reverberation effect in the space domain, providing
insight on the relative positions of the sound source and of the receiver
\cite{a262} \cite{a379}. According to the literature, the reverberation time,
$T_{60}$, is independent of the source to microphone configuration;
in contrast to the RIR, the $T_{60}$ measured in the diffuse sound
field is independent of the source to microphone configuration. This
is important for blindly estimating $T_{60}$ from noisy reverberant
speech \cite{a367}.

The reverberation time, $T_{60}$, is independent of the source to
microphone configuration and depends on the room. The impact of reverberation
on human auditory perception depends on the reverberation time. If
$T_{60}$ is small, the environment reinforces the sound which may
enhance the sound perception \cite{a286}. On the contrary, if $T_{60}$
is large, a spoken syllable may persist for long and interfere with
future spoken syllables.

According to \cite{a285}, dereverberation algorithms that operate
in the power spectral domain are robust and relatively insensitive
to speaker movements and minor variations in the spatial placement
of sources. In this context, algorithms that leave the phase unaltered
and operate in the amplitude, power or log-power spectral domain are
insensitive to speaker movements and to minor variations in the spatial
placement of sources.

Enhancement algorithms that perform reverberation suppression, as
opposed to reverberation cancellation, do not require an estimate
of the RIR. In this report, we focus on enhancement algorithms that
perform reverberation suppression. In addition, we also focus on algorithms
that assume that the early and late reverberant speech components
are independent and aim to suppress the late reverberant speech component.

Dereverberation can be performed using spectral subtraction to remove
reverberant speech energy by cancelling the energy of preceding speech
phonemes in the current time-frame.

In \cite{Habets2007}, spectral enhancement methods based on a time-frequency
gain, originally developed for the purpose of noise suppression, have
been modified and used for dereverberation. Such algorithms suppress
late reverberation assuming that that the early and late reverberation
components are independent. The novelty of the algorithms in \cite{Habets2007}
is that denoising algorithms can be adjusted to operate in noisy and
reverberant conditions. Spectral enhancement dereverberation methods
can be easily implemented in the STFT domain and have low computational
complexity. The spectral enhancement dereverberation methods in \cite{Habets2007}
estimate the late reverberant spectral variance (LRSV) and use it
in the place of the noise spectral variance; these algorithms reduce
the problem of late reverberation suppression to the problem of estimating
the LRSV blindly from reverberant speech observations \cite{Lebart2001}.

The idea that late reverberation can be treated as an additive disturbance
originates from \cite{Lebart2001}. In \cite{Habets2007}, this idea
of treating late reverberation as an additive disturbance is expanded
and utilised in various spectral enhancement dereverberation algorithms.
The late reverberation suppression algorithm in \cite{Lebart2001}
statistically models the RIR in the time domain, estimates the LRSV
and uses spectral subtraction to enhance speech.

The seminal work of \cite{Lebart2001} is discussed in \cite{a286}
where a dereverberation algorithm based on blind spectral weighting
is developed to suppress late reverberation and reduce its overlap-masking
effect. According to \cite{a286}, the late reverberant speech component
causes overlap-masking that smears the high energy phonemes, such
as the vowels, over time, fills envelope gaps and increases the prominence
of low-frequency energy in the speech spectrum. The spectral weighting
algorithm in \cite{a286} mitigates the effect of overlap-masking
using the uncorrelated assumption for late reverberation \cite{Lebart2001}
\cite{Habets2007}.

Estimation of the LRSV is also referred to as reverberation noise estimation.
Several spectral enhancement algorithms that employ different methods
for reverberation noise estimation have been developed in the past.
According to the literature and to \cite{a253}, the LRSV estimator
presented in \cite{a277} is a continuation and an extension of the
LRSV estimator in \cite{Lebart2001}. The dereverberation algorithm
in \cite{a277} statistically models the RIR in the STFT domain, and
not in the time domain as \cite{Lebart2001}. Late reverberation is
estimated and suppressed in \cite{a277} by considering the reverberation
time, $T_{60}$, and the energy contribution of the direct path and
reverberant parts of speech in the STFT domain. The DRR is externally
estimated in \cite{a277}. Two common criticisms of spectral enhancement
algorithms that are based on reverberation noise estimation are that
they introduce musical noise and that they suppress speech onsets,
when they over-estimate the true reverberation noise.

According to the literature and to \cite{a283}, ideal ratio masks
and complex ideal ratio masks have been used by researchers for dereverberation.
Complex ideal ratio masks take account of the speech phase since they
estimate the real and imaginary parts of the complex STFT domain of
clean speech. Complex ideal ratio masks estimate either the real and
imaginary parts or the log real and log imaginary parts of the complex
STFT domain of speech. In particular, complex ideal ratio masks utilise
supervised learning and NNs to estimate either the real and imaginary
parts or the log real and log imaginary parts of the complex STFT
domain of clean speech. The NN-based data-driven speech enhancement
algorithm in \cite{a283} uses complex ideal ratio masks for joint
denoising and dereverberation.

In \cite{a276}, the authors do not agree with the claim that complex
ideal ratio masks can be used for dereverberation. In particular,
the data-driven enhancement algorithm in \cite{a276} performs NN-based
blind dereverberation using the Fourier transform of the STFT of the
reverberant speech signal.

Supervised learning and NNs can be used for joint denoising and dereverberation
that is not based on ideal ratio masks and complex ideal ratio masks.
The NN that is used in the speech enhancement algorithm in \cite{a242}
operates in the log-spectral domain, utilises context frames (i.e.
neighboring frames, past and future frames at every time step) and
estimates clean speech from noisy and reverberant speech in the log-spectral
domain. In \cite{a287}, two supervised dereverberation algorithms
are examined: the one NN-based algorithm predicts speech in the amplitude
spectral domain using direct mapping and the other NN-based algorithm
predicts the ideal ratio mask. According to the results of \cite{a287},
NNs used for ideal ratio masking \cite{a379} outperform NNs used
for predicting the speech spectrum in terms of quality and intelligibility
metrics.

We note that the NN-based data-driven speech enhancement algorithm
in \cite{a242} estimates the clean speech phase using a post-processing
technique. More specifically, the supervised algorithm in \cite{a242}
uses an iterative procedure to reconstruct the time-domain signal
that is based on \cite{a291}, which was published in 1984. According
to \cite{a242} and to \cite{a291}, the enhancement algorithm ``iteratively
updates the phase at each step by replacing it with the phase of the
STFT of its ISTFT'', while keeping the target magnitude from the
NN fixed.

In \cite{a274}, NMF is extended to include reverberation. More specifically,
the two single-channel speech enhancement algorithms that are introduced
in \cite{a274} model the room acoustics using a non-negative approximation
of the convolutive transfer function and model speech in the amplitude
spectral domain using NMF. The two speech enhancement algorithms in
\cite{a274} enhance the quality of speech in noisy and reverberant
conditions. A particular advantage of NMF-based algorithms is the
use of iterative multiplicative update rules. Regarding NMF and dereverberation,
the speech enhancement algorithm in \cite{a275} performs joint denoising
and dereverberation using nonnegative matrix deconvolution and nonnegative
speech dictionary models in the amplitude STFT spectral domain.

The modeling of the speech temporal dynamics can be beneficial in
reverberant conditions \cite{a241}, especially in severe reverberant
conditions where the DRR is low and the $T_{60}$ is long. The enhancement
algorithm in \cite{a241} performs both noise reduction and dereverberation
using state-space modeling and speech and noise tracking. Moreover,
the SPENDRED algorithm \cite{a179} \cite{a175} also considers speech
temporal dynamics.

The SPENDRED algorithm, which is presented in \cite{a179} \cite{a175},
performs time-varying $T_{60}$ and DRR estimation and it internally
(and not externally) estimates $T_{60}$ and DRR at every time step.
However, unless the source or the microphone are moving around, the
$T_{60}$ and DRR will presumably be constant throughout the recording.
In addition, SPENDRED also performs frequency-dependent $T_{60}$
and DRR estimation; according to the ACE challenge \cite{a257}, performing
frequency-dependent $T_{60}$ and DRR estimation is important. Furthermore,
SPENDRED performs intra-frame speech correlation modeling; typical
speech enhancement algorithms do not perform intra-frame frequency
correlation modeling and decouple different frequency dimensions,
treating each frequency bin on its own. Decoupling different frequency
dimensions makes the algorithms easier to implement since frequency
bins can be processed in parallel \cite{a268}. On the contrary, modeling the intra-frame
correlation of the clean speech signal is important in order to enhance
the pitch and the harmonics of speech.

Reverberation is frequency dependent and the SPENDRED algorithm takes
advantage of this observation. Estimating frequency-dependent reverberation
parameters is beneficial. Reverberation is frequency dependent and
obtaining a $T_{60}$ estimate for each individual frequency bin,
or for every Mel-spaced frequency band as in \cite{a179} \cite{a175},
is advantageous.

The SPENDRED dereverberation algorithm is a model-based technique;
it uses the reverberation model that is described by the equations
(9.4) and (9.5) in section 9.2 in \cite{a109}. The SPENDRED algorithm
does not utilise the coarser reverberation model that is described
by the equation (9.6) in \cite{a109}, which approximates the square
of the RIR with its envelope only. As the joint denoising and dereverberation
enhancement algorithm in \cite{a237}, SPENDRED employs a parametric
model of the RIR that is based on white noise with a decaying envelope,
in which the decay time of the envelope is given by $T_{60}$.

The enhancement algorithms described in \cite{a282}, \cite{a237}
and \cite{a181} are based on creating statistical observation models
of noisy and reverberant speech in the logarithmic Mel-power spectral
domain. Observation models are used in the KF update step in modulation-domain
Kalman filtering algorithms. Equations (9) and (10) in \cite{a282}
define the observation model that relates noisy and reverberant speech,
speech, reverberation and noise in the logarithmic Mel-power spectral
domain. More specifically, equation (9) in \cite{a282} defines the
observation model that relates noisy and reverberant speech, reverberant
speech and noise in the logarithmic Mel-power spectral domain and
equation (10) in \cite{a282} defines the observation model that relates
reverberant speech, speech and reverberation in the logarithmic Mel-power
spectral domain. In this context, reverberation in the logarithmic
Mel-power spectral domain refers to finding a representation of the
RIR in the logarithmic Mel-power spectral domain. The algorithm in
\cite{a282} uses the instantaneous reverberant-to-noise ratio and
the observation model in the logarithmic Mel-power spectral domain.

The model-based enhancement algorithms presented in \cite{a282},
\cite{a237} and \cite{a181} are also discussed and explained in
section 9.7.3 in \cite{a109}. Section 9.3 in \cite{a109} examine
the equations that define the relations between reverberant and noisy
speech, reverberant speech, speech, reverberation and noise in different
spectral time-frequency domains. Section 9.7.3 in \cite{a109} discusses
a few model-based dereverberation algorithms that use the equations
that define the relations between reverberant and noisy speech, reverberant
speech, speech, reverberation and noise in a specific spectral time-frequency
domain.

As described and discussed in \cite{a238}, the phase factor in Mel-spaced
frequency bands has a different equation, different properties, a
different distribution and different moments from the phase factor
in STFT bins, $\alpha$. In addition, as described and discussed in
\cite{a282}, \cite{a237} and \cite{a181}, the phase factor between
reverberant speech and noise is different from the phase factor between
speech and noise, $\alpha$. In \cite{a237} and in \cite{a181},
the phase factor between reverberant speech and noise in Mel-spaced
frequency bands is examined, investigated and modeled.

In noisy and reverberant conditions, finding the onset of speech phonemes
and determining which frames are unvoiced/silence is difficult because
reverberation tends to spread speech energy over time. In addition,
in noisy and reverberant conditions, noise estimation is difficult
because unvoiced/silence frames are hard to identify and the noise
estimate is affected by the reverberation present in the noisy reverberant
signal. According to \cite{a241}, it is not efficient for enhancement
algorithms to perform a two step procedure that is comprised of a
denoising stage followed by a dereverberation step. The concatenation
of different techniques for noise reduction and dereverberation is
inefficient because denoising and blind dereverberation are not performed
jointly \cite{a241}.

Despite the claim that it is not efficient for algorithms to perform
a two step procedure that is comprised of a denoising stage followed
by a dereverberation step, long-term linear prediction with pre-denoising
can be utilised to suppress noise and late reverberation. According
to the literature \cite{a254} \cite{a109}, with long-term linear
prediction, the effect of reverberation may be represented as a one-dimensional
convolution in each frequency bin. The convolutive nature of reverberation
induces a long-term correlation between a current observation and
past observations of reverberant speech \cite{a249} and this long-term
correlation can be exploited to suppress reverberation. According
to \cite{a248} \cite{a249}, long-term linear prediction using the
weighted prediction error (WPE) algorithm can be utilised for late
reverberation reduction and is robust to noise. In \cite{a250}, long-term
linear prediction is discussed along with inter-frame speech correlation
modeling. The algorithm in \cite{a248} utilises the WPE algorithm
and long-term linear prediction in the complex STFT domain. According
to \cite{a249}, the WPE algorithm can also be used in the power spectral
domain: the algorithm in \cite{a249} examines the possibility of
subtracting the power spectra of the reverberation estimates from
the observed power spectra while leaving the phase unchanged instead
of subtracting the reverberation estimates in the STFT domain.

The speech enhancement algorithm in \cite{a284} perform a two step
procedure that is comprised of a denoising stage followed by a dereverberation
step. In \cite{a284}, NN-based pre-denoising is used; the dereverberation
step is performed using the WPE algorithm. Figure 1.b in \cite{a284}
and Sec. 4 in \cite{a284} describe the NN, which operates in the
log-spectral domain, that is used for pre-cleaning the noisy and reverberant
speech signal. The pre-cleaned power spectral domain is then used
for the WPE algorithm; a particular feature of the algorithm in \cite{a284}
is that the WPE method does not need more than one iterations.

The WPE linear filtering approach removes reverberation in the complex
STFT domain taking consecutive reverberant observations into account
\cite{a109}. An adaptive and multi-channel variant of the WPE algorithm
has recently been used as a front-end dereverberation method in ``Google
Home'' \cite{a267}.

According to the literature and to \cite{a249} and \cite{a254},
for dereverberation, linear filtering can either exploit both the
spectral amplitudes and phases of the signal or exploit the spectral
amplitudes and leave the spectral phase unaltered. The speech spectral
phase is severely affected by reverberation because reverberation
is a superposition of numerous time-shifted and attenuated versions
of the clean speech signal. It is worth noting that reverberation
is strongly correlated with clean speech both in the short-term and
in the long-term.

According to \cite{a248}, we can estimate reverberation in the complex
STFT domain, $R_t$, performing a few iterations and using $T_1=3$
and $T_2=40$. The parameter $T$ is the number of frames of the entire
speech utterance; WPE performs batch processing and operates on the
entire speech utterance.

The WPE method is an iterative algorithm that alternatively estimates
the reverberation prediction coefficients and the speech spectral
variance using batch processing of speech utterances. The WPE method
needs the entire speech utterance for processing. Therefore, one of
the drawbacks of the WPE method is that it requires at least a few
seconds of the observed speech utterance in order to ensure the convergence
of the reverberation prediction coefficients \cite{a254}. In addition,
it is worth noting that the RIR should remain constant \cite{a254}.

According to \cite{a253}, using WPE to estimate the reverberant component
of speech leads to a processing delay. The WPE method is a batch processing
technique and it requires the pre-processing of the entire speech
utterance in order to provide an accurate estimate of the reverberant
component of speech \cite{a253}. Batch processing is not suitable
when dealing with time-varying acoustic environments with varying
RIRs. In \cite{a253}, WPE is utilised for processing non-overlapping
blocks of $0.5$ s long. Equations (15)-(19) in \cite{a253} describe
the block-wise WPE method that can be used in real-world environments.

In summary, in this literature review, several different enhancement
algorithms for noise suppression and dereverberation were presented,
explained and discussed. One of the main points is that different
enhancement algorithms operate in different spectral time-frequency
domains and follow different methodologies and frameworks. Speech
is a non-white signal and its correlation structure should not be
destroyed; the speech enhancement algorithm needs to be able to distinguish
between the correlation introduced by the RIR and the correlation
of the speech signal itself \cite{a109}. A final remark is that real-world
speech recordings are inevitably distorted by both noise and frequency-dependent
reverberation \cite{a289} \cite{a379}.

\section{Conclusion }

This report focuses on speech enhancement considering both noise and
convolutive distortions \cite{a172} \cite{a379}. Additive noise and room reverberation
are two different types of distortion and the effects of both need
to be suppressed and ideally eliminated \cite{a379}. The effects of additive noise
are limited to a single frame of short-time signal analysis while
the effects of room reverberation span a number of consecutive time
frames. Non-linear adaptive modulation-domain Kalman filtering algorithms
can be used for speech enhancement, i.e. noise suppression and dereverberation,
as in \cite{a367}, \cite{a268}, \cite{a225}, \cite{a360} and \cite{a236}.
Modulation-domain Kalman filtering can be applied for both noise and
late reverberation suppression; in \cite{a268}, \cite{a367}, \cite{a225}
and \cite{a360}, various model-based speech enhancement algorithms
that perform modulation-domain Kalman filtering are designed, implemented
and tested. The model-based speech enhancement algorithm presented
in \cite{a268} tracks and estimates the clean speech phase and the
STFT-based algorithm described in \cite{a236} uses the active speech
level estimator presented in \cite{a102}.

\selectlanguage{british}%
\bibliographystyle{IEEEtran}
\addcontentsline{toc}{section}{\refname}\bibliography{sapref}
\selectlanguage{english}%

\end{document}